\documentclass[fleqn,10pt]{wlscirep}
\usepackage[utf8]{inputenc}
\usepackage[T1]{fontenc}
\usepackage[normalem]{ulem}
\usepackage{multirow}
\usepackage{silence}
\usepackage{lineno}
\title{
Halide Perovskite Light-Emitting Photodetector \\
}
\author[1,*]{A.A.Marunchenko}
\author[1]{V.I.Kondratiev}
\author[1]{A.P.Pushkarev}
\author[1]{S.A.Khubezhov}
\author[1]{M.A.Baranov}
\author[2]{A.G.Nasibulin}
\author[1,3,4*]{S.V.Makarov}

\affil[1]{ITMO University, School of Physics and Engineering, St. Petersburg, 197101, Russian Federation}
\affil[2]{Skolkovo Institute of Science and Technology, 30/1 Bolshoy Boulevard, 121205 Moscow, Russian Federation}
\affil[3]{Harbin Engineering University, Harbin 150001, Heilongjiang, China}
\affil[4]{Qingdao Innovation and Development Center of Harbin Engineering University, Qingdao 266000, Shandong, China}

\affil[*]{Corresponding author}

\makeatletter
\renewcommand{\@maketitle}{%
{%
\thispagestyle{empty}%
\vskip-36pt%
{\raggedright\sffamily\bfseries\fontsize{20}{25}\selectfont \@title\par}%
\vskip10pt
{\raggedright\sffamily\fontsize{12}{16}\selectfont  \@author\par}
\vskip25pt%
}%
}%
\makeatother

\begin{document}

\flushbottom
\maketitle

\section*{ABSTRACT}

Light emission and detection are the two fundamental attributes of optoelectronic communication systems. Until now, both functions have been demonstrated using the p-n diode which is exploited across a wide range of applications \cite{sze2021physics}. However, due to the competing dynamics of carrier injection and photocarrier collection \cite{rau2007reciprocity}, with this device light emission and detection are realized separately by switching the direction of the applied electrical bias \cite{bie2017mote,bao2020bidirectional,oh2017double}. Here we use mobile ions in halide perovskites to demonstrate light-emitting photodetection in either condition of applied electrical bias. Our device consists of a CsPbBr$_3$ microwire which is integrated with single-walled carbon nanotube thin film electrodes. The dual functionality stems from the modulation of an energetic barrier caused by the cooperative action of mobile ions with the photogenerated charge carriers at the perovskite-electrode interface. Furthermore, such complex charge dynamics also result in a novel effect: light-enhanced electroluminescence. The observed new optoelectronic phenomena in our simple lateral device design will expand the applications for mixed ionic-electronic conductors in multifunctional optoelectronic devices \cite{lee2020multifunctional}.

\hfill \break


\thispagestyle{empty}

\par The multifunctionality in optoelectronics supposes several physical phenomena involved during the device operation to serve the application demands \cite{lee2020multifunctional}. Intuitively, its possible realization is the merging of functions from separate devices combined into one architecture \cite{niu2019wireless}. However, this approach requires accurate preparation of the parts and their matching. Another approach is to optimize device operation through a controllable external stimulation. A good example is the p-n junction which can operate as a light emitter at forward bias direction and as a photodetector in the reverse bias mode \cite{bie2017mote,bao2020bidirectional,oh2017double}. 
Alternatively, multifunctionality can be achieved by introducing new responsive materials to the device possessing a certain function. For instance, the exposure of p-AlGaN/n-GaN junction to electrolyte environment serving as an additional electrode enables light wavelength dependent polarity switching of photocurrent \cite{wang2021bidirectional}. 

\par Along with the aforementioned approaches, there is significant interest to materials which already have structural properties affording simultaneous multiple responses. Halide perovskites are among of such materials, which in addition to their outstanding optoelectronic properties (long carrier diffusion, high carrier mobility, defect tolerance and large carrier absorption coefficient) \cite{green2014emergence, herz2017charge, stranks2013electron}, exhibit intrinsically complex mixed ionic-electronic conduction \cite{eames2015ionic}. In fact, the additive mobile ionic species in halide perovskites dynamically respond to various applied stimuli such as light \cite{kim2018large}, temperature \cite{bag2015kinetics, zhang2020defect} and electric field \cite{xiao2015giant,li2018unravelling}. Their significant impact on the device performance can be separated from electronic charge carriers \cite{kim2018large}. Indeed, the halide vacancy migration driven by applied bias modulates the local perovskite photoluminescence intensity \cite{li2018unravelling}. Furthermore, the applied electric field of 1 V/$\mu$m causes ionic redistribution in bulk perovskite leading to a switchable photovoltaic effect in metal-perovskite-conductive oxide design \cite{xiao2015giant}. 

\par The sensitivity of mobile ions in halide perovskites to various stimuli is generally recognized to be the main obstacle to stable working perovskite-based common optoelectronic devices, such as solar cells (SCs), photodetectors (PDs) or light-emitting diodes (LEDs) \cite{liu2021correlations, yuan2016ion}. In these devices operating at either DC bias or built-in potential, the electric field-affected mobile ions deteriorate performance characteristics \cite{liu2021correlations}. Despite that, some emerging state-of-the-art applications avoiding light stimuli such as ionic memories and switchers take advantage of perovskite mobile ions \cite{yen2021all, zhumekenov2021stimuli}. However, the controllable ionic migration in halide perovskites upon light illumination and applying electric field has not been employed for truly multifunctional optoelectronic devices. From this point of view, unique structural properties of light-sensitive ionic semiconductors can be exploited in a rational way.

\par In this work we present a strategy to create a light-emitting photodetector operating at given electrical bias. The dual functionality (photodetection mode, light-emission mode) is achieved due to migration of perovskite ions modulating energetic barriers at the interfaces in the cooperation with the photogenerated carriers. The device is as simple as cesium-lead tribromide (CsPbBr$_3$) microwire (MW) connected to single-walled carbon nanotube (SWCNT) thin film electrodes. We demonstrate the operation of light-emitting photodetector at room temperature featuring the light-enhanced electroluminescence property . 

\begin{figure}[t!]
\centering
\center{\includegraphics[width=1.0\linewidth]{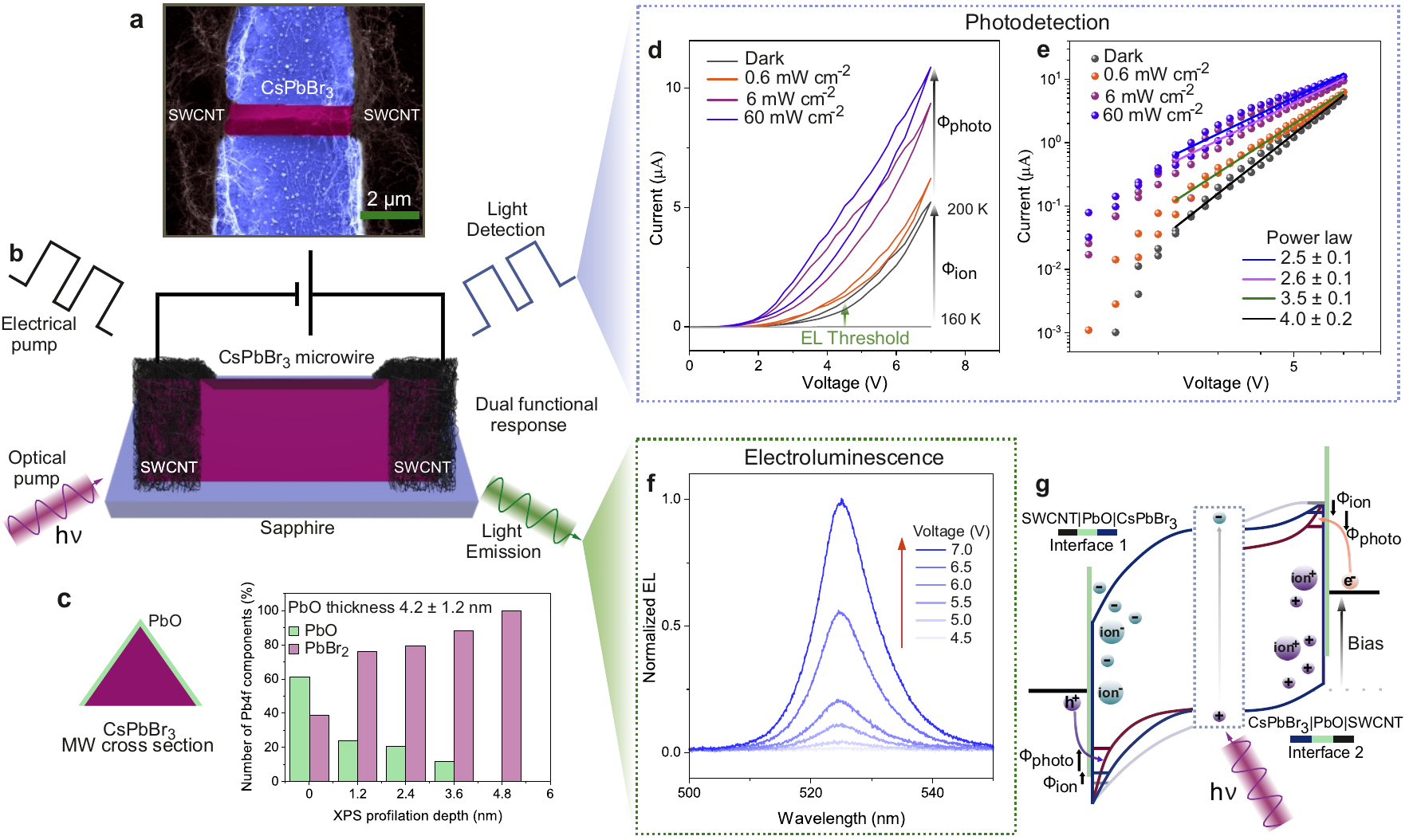}}
\caption{ | \textbf{The structure, design and operation of the CsPbBr$_3$ microwire device at 200 K}. 
\\ \textbf{a} SEM image of the fabricated device: a single CsPbBr$_3$ microwire (MW) connects two SWCNT film electrodes at a distance of $\sim$ 5 $\mu m$. \textbf{b} The scheme of the device operation where the complex optical and electrical pump result in the optoelectronic dual functional response at a given polarity of applied electrical bias. \textbf{c} The schematic cross-section of MW with lead oxide (PbO) layer. The XPS depth profiling reveals thickness of PbO 4.2 $\pm$ 1.2 nm. \textbf{d,e} Current-voltage curves for the device in dark and illuminated conditions. Dark current at 200 K dramatically increases as compared to one at 160 K. The EL threshold is revealed at 4.5 V. The photocurrent is collected in the 0.6 $-$ 60 mW cm$^{-2}$ intensity range at 490 nm wavelength CW illumination. The linear fits \textbf{e} are done in range 2.25 $-$ 7 V and reflect gradual slope decrease from 4.0 $\pm$ 0.2 to 2.5 $\pm$ 0.1 with the increase in illumination. \textbf{f} Normalized EL spectra at 4.5 $-$ 7 V applied bias. \textbf{g} Schematic band diagram of the device under applied bias where the interface regions SWCNT|PbO|CsPbBr$_3$ are highlighted. The effective energetic barrier height for the carrier injection is modulated by both mobile ions $\Phi_{ion}$ and photocarriers $\Phi_{photo}$. The colors reflect the band diagram under different conditions: grey - without charge carriers, blue - with mobile ions, and red - with mobile ions and photocarriers. The corresponding red and blue horizontal lines reflect the position of the barrier for the charge carrier (electrons or holes) injection. 
}
\end{figure}

\par CsPbBr$_3$ perovskite MWs are fabricated  by  temperature difference triggered growth method (for details, see Methods) \cite{wang2018temperature}. The deposition of perovskite is conducted on sapphire substrates affording the MWs directional growth (Supplementary Fig.S1) and excessive heat collection during the device operation owing to high thermal conductivity of sapphire. The typical photoluminescence spectrum collected from an ensemble of MWs is peaked at 521 nm and demonstrates full width at half maximum $\delta\lambda$ = 17.8 nm (Supplementary Fig.S1). X-ray diffraction pattern (Supplementary Fig.S2) proves high crystallinity of as-synthesized microwires possessing an orthorhombic structure (space group P$_{nma}$) \cite{saidaminov2017inorganic}. A stoichiometric atomic ratio of Cs:Pb:Br elements is checked out by the energy-dispersive x-ray analysis and equals to 1:1:3 (Supplementary Fig.S3).

\par Afterwards, SWCNT thin film synthesized by CVD method (for details, see Methods) is dry-transferred on top of the as-grown CsPbBr$_3$ MWs. Then, precise femtosecond laser ablation of the film is performed.  Generally, 1030 nm fs laser light focused at 3 $\mu$m spot with fluence below 0.1 J/cm$^2$ and 100 Hz repetition rate results in gentle removing of SWCNT film \cite{marunchenko2022single} which does not affect the surface morphology of MWs and gives an interelectrode distance of $\sim$ 5 $\mu$m (Fig.1a, Supplementary Fig.S4). The electrode function of SWCNT film is complemented by its chemical inertness which increases the operational stability of perovskite-based optoelectronic devices \cite{marunchenko2022single,aitola2017high}.

\par Eventually, the as-fabricated device is a single CsPbBr$_3$ MW short-circuiting two symmetrical laser ablated SWCNT film electrodes (Fig.1a,b, Supplementary Fig.S4). In context of semiconductor device physics, it should be referred to metal-semiconductor-metal MSM structure \cite{sze2021physics, sharma2013metal}. In general, the metal-semiconductor interface may dramatically influence the device performance. The fabrication of our device in the presence of atmosphere can lead to the formation of native metal oxide. Therefore, we perform X-ray photoelectron spectroscopy profiling (Fig.1c) which proves the formation of lead(II) oxide (PbO) of 4.2 $\pm$ 1.2 nm thickness (Supplementary Fig.S5,S6,S7). The PbO should affect carrier transport in our MSM Schottky barrier-type device by introducing an additional energetic barrier for selective charge transport \cite{sze2021physics, sharma2013metal}.

Charge carrier transport in the fabricated device is governed by coupled drift-diffusion and Poisson equations \cite{moia2019ionic,bertoluzzi2020mobile}. The latter can be written for CsPbBr$_3$ MW at coordinate $x$ as:
\begin{equation}
\frac{\partial^2{\phi}}{\partial {x^2}} = -\frac{q}{\epsilon_0\epsilon_r} (N^{+} - N^{-}),  
\end{equation}
where $\phi$ is electrostatic potential, $x$ position of charges in one-dimension, $q$ is electron charge, $\epsilon_0$ and $\epsilon_r$ are the dielectric constant of vacuum and perovskite material relative permittivity, respectively. Finally, $N^{+}$ and $N^{-}$ are concentrations of positive and negative charges in perovskite. It is worth noting that the right part of \textbf{eq.1} does not reflect the nature of charge carrier type. This means that all of the present charge carriers (both ionic and electronic) govern the electric potential and electric field at both SWCNT|CsPbBr$_3$ and CsPbBr$_3$|SWCNT symmetrical interfaces. 
\par For the case of metal-semiconductor (MS) interface, consider for example electron transport. The accumulation of positive carriers with concentration N$^{+}$ near MS interface region with band bending \emph{V$_{bb}$} enhances the electric field E and causes Schottky barrier $\Delta\Phi$ lowering \cite{tung2001recent,sze2021physics}:
\begin{equation}
E = \sqrt{\frac{2 q  N^+ V_{bb}}{{\epsilon_0}{\epsilon_r}}}
\end{equation}
\begin{equation}
\Delta\Phi = \sqrt[4]{\frac{2 q ^3  N^+ V_{bb}}{8 \pi^2 {\epsilon_0}^2 {\epsilon_r}^2}}
\end{equation}
This way, positive charges near MS interface N$^{+}$ lead to more efficient thermionic emission of electrons as well as enhance the probability of the tunneling through the MS interface, which leads to current increase \cite{padovani1966field,crowell1969normalized,tung2014physics}. 
In our system, hybrid optoelectronic pump (Fig.1b) gives photogenerated holes and positively charged mobile ions attributing to carrier $N^{+}$. The opposite is true for the hole transport through the SWCNT|CsPbBr$_3$ interface, where similar negative charge species contribute to $N^{-}$. We further refer to the role of accumulating species ($N^{+}$,$N^{-}$) which affect carrier injection at MS interface as effective energetic barrier lowering, emphasizing mixed thermionic-field emission behaviour \cite{padovani1966field,crowell1969normalized,tung2014physics}.

\par The DC current-voltage (I-V) scans are performed in the 40 $-$ 200 K temperature range. This allows us to separate photogenerated carriers from mobile ions which possess temperature dependent transport \cite{bag2015kinetics}. The nonlinearity of I-V curves identifies the Schottky-type contacts at SWCNT|CsPbBr$_3$ interfaces (Fig.1d,e, Supplementary Fig.S8,S9). Remarkable behavior is observed when elevating temperature from 160 K to 200 K where the dark current magnitude drastically changes up to 1.8 $\cdot$ 10$^4$ times (Fig.1d, Supplementary Fig.S8). This temperature range is identified as characteristic for the mobilisation of CsPbBr$_3$ ionic species \cite{zhang2020defect}. Another evidence of the ionic motion at 200 K is the sudden change in dark current hysteresis behaviour as compared to that of observed at temperature below 160 K (Supplementary Fig.S8) \cite{lenes2011operating, yen2021all}. The mobile ions play role in effective energetic barrier lowering owing to their accumulation at SWCNT|CsPbBr$_3$ interface according to \textbf{eq.1-3}. Therefore, it is constructive to consider the effective energetic barrier modulation $\Phi_{ion}$ which gates the electronic-type current at temperatures sufficient to make the perovskite ions mobile. 

\par The photocurrent invoked by continuous-wave (CW) illumination with intensity of 0.6$-$60 mW cm$^{-2}$ at wavelength 490 nm is detected as the gradual current increase (Fig.1d,1e,Supplementary Fig.S8,S9). Thus our device exhibits photodetection (PD). Notably, in the device illuminated at 200 K, not only mobile ions modulate the current via $\Phi_{ion}$, but also photogenerated carriers via $\Phi_{photo}$ do \textbf{eq.1-3}. For this case, the log($I$)/log($V$) plot in dark and light conditions is shown in (Fig. 1e) and the power law $V$ $\sim$ $I^k$ ($k$ is positive integer) is used to fit the data in the 2.25$-$7 V range. In dark conditions, the slope 4.0 $\pm$ 0.2 identifies the trap-filling regime (k > 2), whereas upon light illumination it decreases down to 2.5 $\pm$ 0.1 at 60 mW cm$^{-2}$ approaching space-charge-limited regime (k = 2) \cite{zhang2021space}. 

\par In addition to photodetection, we separately reveal electroluminescence (EL) from not illuminated single CsPbBr$_3$ MW at bias above 4.5 V (Fig.1f). The appearance of EL at 200 K means the mobile ions modify the electronic component of current to a sufficient extent for making radiative recombination to be observable. The $\delta\lambda$ value of 11 nm and the peak wavelength 525 nm are achieved at 7 V and 200 K.

\begin{figure}[t!]
\centering
\center{\includegraphics[width=0.98\linewidth]{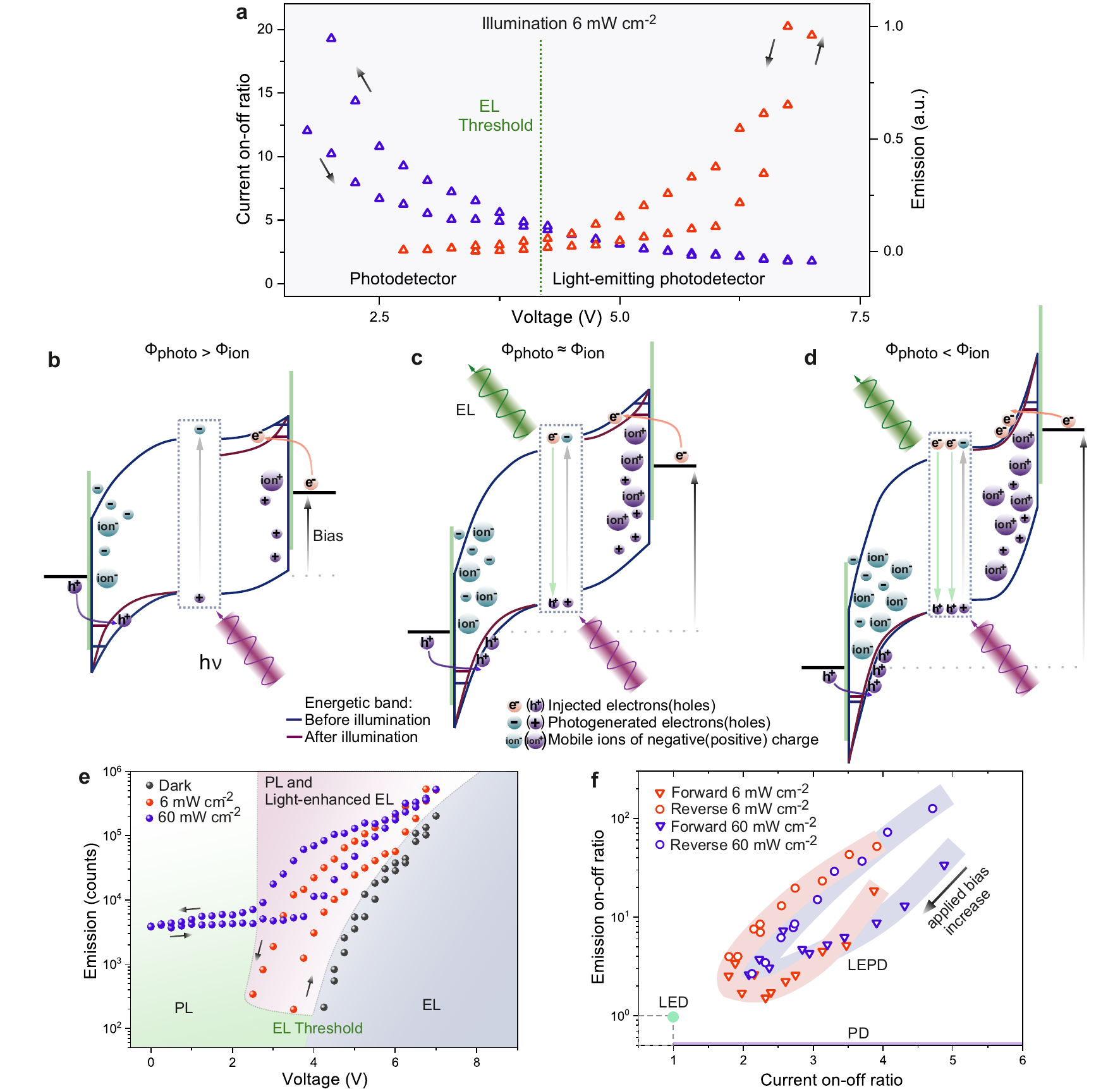}}
\caption{ | \textbf{LEPD operation at DC at 200 K}. 
\\ \textbf{a} Current on-off ratio with the integrated emission versus voltage upon 6 mW cm$^{-2}$ illumination intensity. The EL threshold separates photodetection (PD) from light-emitting photodetection (LEPD) regimes. Arrows indicate the voltage scan direction. \textbf{b,c,d} The device operation at different applied bias relevant to PD, LEPD regime and the transition between them in terms of energetic bands at interfaces SWCNT|PbO|CsPbBr$_3$. The relative concentration of charges (ions and photocarriers) at the interfaces define the relation between $\Phi_{ion}$ and $\Phi_{photo}$ gates. Electrical pump (bias) and optical pump ($h\nu$) result in LEPD causing increase in current (PD) as well as EL appearance. Blue color reflects band diagram in dark, and red upon illumination. \textbf{b} At low applied bias, the modulation $\Phi_{photo}$ is relatively high, but EL is absent. \textbf{c} Near EL threshold the operation of LEPD is defined by balanced $\Phi_{ion}$ and $\Phi_{photo}$. \textbf{d} EL of LEPD is enhanced while the relative $\Phi_{photo}$ decreases reflected in the current on-off ratio decline \textbf{a}. \textbf{e} Three regimes are separated when LEPD is illuminated: (i) EL - in dark; (ii) PL in illuminated case, when dark currrent is low; (iii) combination of PL and light-enhanced EL. \textbf{f} The novel figure of merit for LEPD shows both current and emission on-off ratio upon the illumination. The LEPD is compared to conventional p-n junction LED and PD devices. 
}
\end{figure}

\par To clarify visualisation of $\Phi_{ion}$ and $\Phi_{photo}$ gating phenomena, we present the energetic bands of the device during operation (Fig.1g). The SWCNT work function (5.0 eV) as well as conduction band minimum (electron affinity) and valence band maximum for PbO (3.4 eV and 6.2 eV) and CsPbBr$_3$ (4.2 and 6.5 eV) layers are taken from previous reports \cite{shiraishi2001work,liao2016difficulty,tao2019absolute}. Here, CsPbBr$_3$ is considered to be lightly p-doped. The mobile ions together with photogenerated carriers change effective energetic barrier via $\Phi_{ion}$ and $\Phi_{photo}$. These effective enegetic barrier lowering is controlled by complex optoelectronic stimuli which result in the device multifunctionality.


\par The discovery of separate PD and EL phenomena in one device at similar bias grounds the search towards intermediate operation regime where they exist simultaneously. We introduce a term light-emitting photodetector (LEPD) for a device which enables this regime. The LEPD regime can be vividly determined by synchronizing electrical pump with a spectrometer, the information on EL at every point of the current-voltage scans is obtained (for details, see Methods). The transition between the pure PD and LEPD occurs at threshold bias for the electroluminescence (Fig.2a). Here we plot current on-off ratio and integrated light emission at 6 mW cm$^{-2}$ illumination intensity versus applied bias. The current on-off ratio is a characteristic of LEPD which separates purely ionic and mixed ionic-photo gating phenomena (Supplementary Fig.S10). 

\par In LEPD regime $\Phi_{ion}$ and $\Phi_{photo}$ cooperatively affect the effective energetic barrier for carrier injection at the interfaces. However, since the gates are defined by accumulative charge $N^{+}$($N^{-}$) (eq.2-3), they compete in terms of their relative contribution to the current $I$ enhancement (Supplementary Fig.S10). We can introduce effective energetic barrier $\Phi(V)$ and consider $I$ $\sim$ $e^{-\Phi(V)}$ similar to ideal diode equation with effective energetic barrier $\Phi(V)$ \cite{sze2021physics,sze1971current,tung2014physics}. For the case of DC scans with increasing voltage (Fig. 2a), the relative concentration of photo- to ionic carriers at interfaces (Fig.2b-d) determine the relation $\Phi_{photo}$ /$\Phi_{ion}$. 
 
\par From the point of view of band-bending, the PD, LEPD regimes and transition between them can be depicted by three diagrams showing all types of charge carriers involved to $\Phi_{ion}$ and $\Phi_{photo}$ gatings during the device operation (Fig. 2b-d) in the 1.5-7.0 V range of applied bias. At a given $\Phi_{photo}$, the $\Phi_{ion}$ increases (under bias and time) enhancing total light emission, but affecting current on-off ratio by decrease of effective energetic barrier change $\Phi(V)$ under $\Phi_{photo}$. At lower voltages PD regime is present with maximum current on-off ratio (Fig. 2a,b). Here EL is absent due to subbandgap voltages and low $\Phi_{ion}$ (Fig. 2e). In contrast, at high voltages relative $\Phi_{photo}$ gate is reduced leading to current on-off ratio decrease down to 1.8 (Fig. 2a,2d). This high-voltage LEPD regime yields EL enhanced by strong barrier modulation because of perovskite mobilie ions $\Phi_{ion}$ (Fig. 2g). Eventually, there is an intermediate regime, in between, where $\Phi_{ion}$ and $\Phi_{photo}$ gates have comparable values $\Phi_{photo}\sim\Phi_{ion}$ so that current on-off ratio is mediate, and EL threshold is overcome (Fig. 2a,2c). 

\par A remarkable feature of the device operating in LEPD regime is improvement of EL intensity in the presence of incident light (Fig. 2e). The feature manifests in the reduction of turn-on voltage from 4.5 V  to 2.75 V (Fig. 2e). This means that in LEPD regime, both stimuli, namely applied voltage and incident light simultaneously enhance radiative recombination via current increase invoked by $\Phi_{ion}$ and $\Phi_{photo}$. At first sight, it is reasonable to assume that photoexcited PL makes a major contribution to the established enhancement since it is not straightforward to separate PL and EL signals (Supplementary Fig.S11). However, it should be noted that PL intensity upon 6 mW cm$^{-2}$ illumination is an order of magnitude lower than EL intensity at turn-on voltage. Moreover, a ten-fold increase in intensity of the incident light up to 60 mW cm$^{-2}$ does not affect the enhancement. Thus, we conclude that despite both mechanisms may occur, the light-enhanced EL is supposed to be main contributor to the emission improvement. 

\par The emergence of new-type devices requires a reliable criteria for the comparison of their performance. As one of such criteria, we suggest figure of merit (FoM) showing emission on-off ratio versus current on-off ratio upon illumination and above EL threshold (Fig. 2f). The proposed FoM reflects the difference between the operational ranges for conventional devices (LED,PD) and LEPD. According to this, a high performance LEPD should demonstrate maximum on-off ratios at minimum applied stimuli (bias, illumination). In (Fig. 2f) one can see that on-off ratios decrease with the increase in applied voltage. Such behavior is explained by recombination kinetics in halide perovskites. In detail, when numerous trap states are present at the interfaces, they cause the accumulation of photogenerated charge carriers and contribute to $\Phi_{photo}$. At low bias, this contribution is the highest because of unfilled trap states. The latter is confirmed by the slope of log-log plot of EL intensity versus current which equals 2 and corresponds to trap-filling dominant recombination regime (Supplementary Fig.S12).


\par The temporary stable response in LEPD requires stationary $\Phi(V)_{ion}$, however dynamic accumulation of ions at the interfaces results in time-dependent $\Phi(V,t)_{ion}$ and is a major source of device instability at DC bias. A similar statement is true for other perovskite-based p-n junction-type devices \cite{eames2015ionic,liu2017lab,yuan2016ion}. Taking this into account, after the discovery of the dual functionality in the DC-biased device which additionally undergoes overheating because of Joule losses, we consider pulsed biasing affording quasi-stationary $<\Phi(V)_{ion}>$ for the device operation at room temperature. Furthermore, such AC biasing helps to resolve LEPD regime in time (Fig.3a). 

\begin{figure}[t!]
\centering
\center{\includegraphics[width=0.99\linewidth]{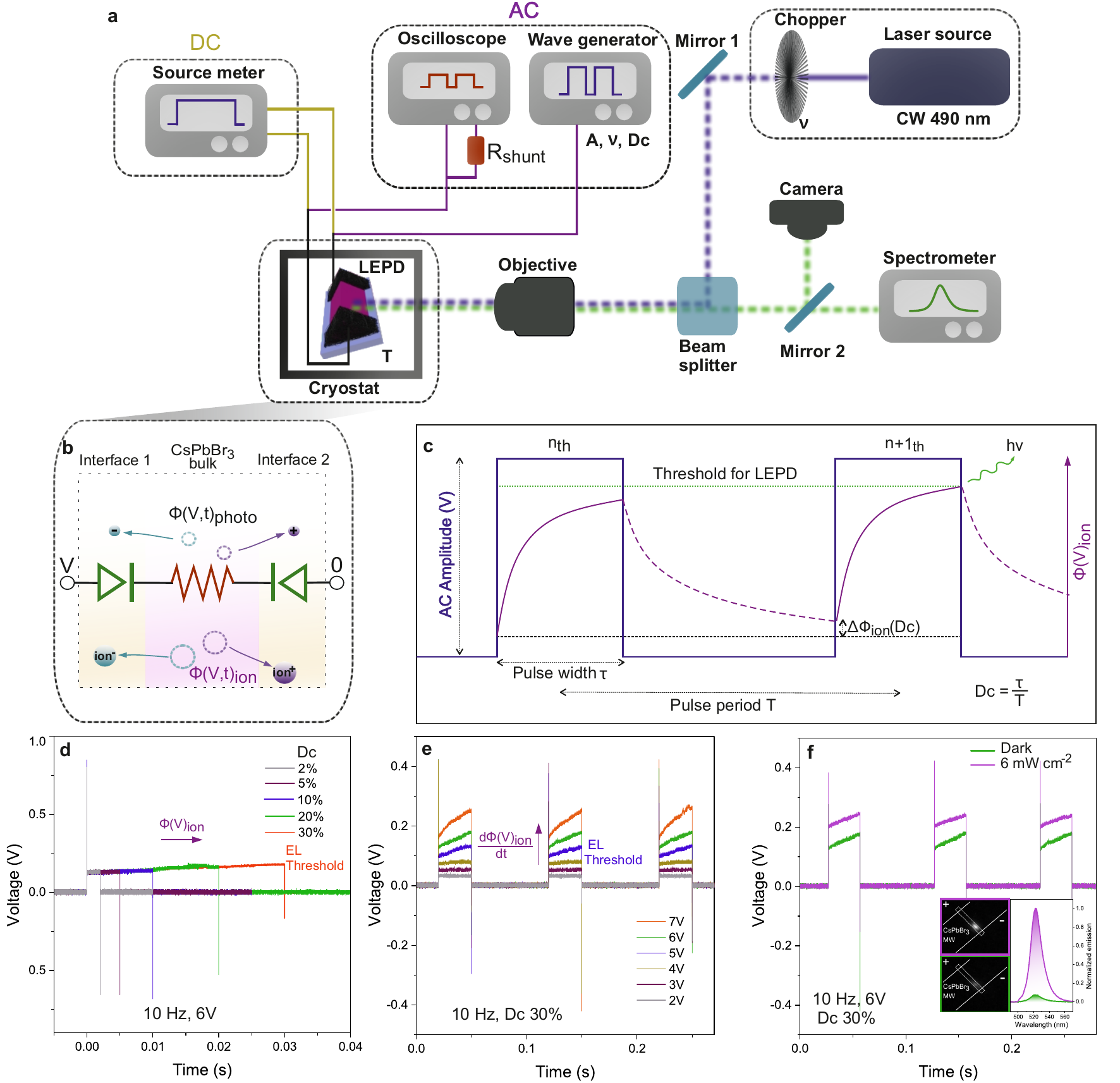}}
\caption{ | \textbf{LEPD operation at AC at room temperature}. 
\\ \textbf{a} Scheme of the experimental setup which is used during device operation. At room temperature AC regime is provided by wave generator of varied voltage amplitude A, pulse frequency $\nu$ and Duty cycle (Dc). The CW laser source is chopped with corresponding optical frequency $\nu$. The optoelectronic signal is measured by oscilloscope and spectrometer. \textbf{b} Equivalent scheme of CsPbBr$_3$ MW presented as two Schottky diodes reflecting the interfaces modulated by time-dependent $\Phi(V,t)_{ion}$ and $\Phi(V,t)_{photo}$. The bulk of CsPbBr$_3$ perovskite provides the interfaces with mobile ions and photocarriers. \textbf{c} The transition to LEPD regime under AC rectanglur pulses is visualised at $n_{th}$ and $n+1_{th}$ pulse. 
Here $\Phi(V)_{ion}$ dynamically increases in time with Dc at fixed pulse frequency. At the end of the pulse $\Phi(V)_{ion}$ relaxation is plotted with dotted curve. The difference $\Delta\Phi_{ion}(Dc)$ is Dc dependent change of $\Phi_{ion}$ between the pulses. \textbf{d-f} The 10 Hz, 6V and Dc 30 $\%$ chosen as optimal for LEPD demonstration at AC \textbf{f}. \textbf{d} $\Phi(V)_{ion}$ is changed during the pulse length leading to EL threshold at 30$\%$ Dc. \textbf{e} The dependence of rate of $\Phi(V)_{ion}$ versus V is demonstrated and the EL threshold appears at 5 V pulse amplitude. \textbf{f} The LEPD operation upon 6 mW cm$^{-2}$ illumination. The current on-off ratio 1.5 with peak emission on-off ratio of 8 (inset image) is achieved. The MW is visualised with its emission area position and the bias polarity (inset image).
}
\end{figure}

\par The LEPD MSM device can be considered as two back to back Schottky diodes which can be modulated and are connected through the perovskite resistive element (Fig.3b). Each of the Schottky diode is the interface between metal (SWCNT film) and semiconductor (CsPbBr$_3$) including PbO layer \cite{sze2021physics, tung2014physics}. Under electrical and optical stimuli, the bulk CsPbBr$_3$ provides the Schottky diodes with photo- and ionic carriers and modulates the total current via $\Phi_{photo}$ and $\Phi_{ion}$, respectively. The rate of the modulation is then defined by drift of these carriers from bulk towards the interface \cite{wang2019kinetic,bernards2007steady}.

\par For the AC biasing, periodic rectangular pulse train with zero bias between the pulses is employed (Fig. 3a,3c),(for details, see Methods). The stimulus is defined by three parameters: voltage amplitude $V$, pulse repetition frequency $\nu$ = $1/T$ and pulse width $\tau$. We schematically illustrate biasing conditions impact on the $\Phi(V)_{ion}$ (Fig. 3c). During the $n_{th}$ pulse of width $\tau$ and voltage amplitude $V$, ions modulate $\Phi(V)_{ion}$. In this case, the time response of LEPD is defined by the speed of reaching $\Phi(V)_{ion}$ sufficient for EL governed by a single-pulse width $\tau$. We identify this operation mode as a single-pulse one. In general, this value may not exceed the limit for LEPD at this pulse width $\tau$ (Fig. 3c). At the end of the pulse, where voltage is not applied, ions relax to their initial position during the timescale of $T - \tau$ which is reflected in $\Phi(V)_{ion}$ decrease (Fig. 3c). At the start of the next ${n+1}_{th}$ pulse, ions may be not completely relaxed during T - $\tau$ interval leading to small $\Delta\Phi_{ion}$ as compared to previous pulse. The ${n+1}_{th}$ pulse reaches the limit for EL threshold and, thus, LEPD regime. Here, LEPD works in accumulative mode and its time response is defined by ${n+1}_{th}$ pulse. We stress that entering LEPD regime can be realized in wide range of pulse repetition frequency, but for a given frequency, the Duty cycle (Dc) = $\nu\tau$ should be chosen according to ionic relaxation dynamics.

\par In (Fig. 3d) the dependence of the current flowing through the series shunt resistance on Dc at frequency 10 Hz and amplitude 6 V is shown. Starting from 2 {$\%$} the Dc is increased up to 30 {$\%$} until the observation of EL. The speed of the ionic modulation $\frac{d\Phi(V)_{ion}}{dt}$ at 30 $\%$ Dc increases with the applied voltage amplitude (Fig. 3e). The EL threshold is achieved at 5 V. In principle, EL is also observed at higher frequencies (e.g. 1 kHz and 100 kHz) relevant to accumulative mode of LEPD (Supplementary Fig.S13). Finally, for the set of AC biasing parameters (10 Hz, 6 V, Dc $\%$) the LEPD operating upon 6 mW cm$^{-2}$ illumination intensity is demonstrated. The photodetection is observed with on-off ratio 1.5 together with light-enhanced electroluminescence with peak emission on-off of 8 (Fig. 3f). This result states the principal ability of both current (PD) and emission (Light-enhanced EL) enhancement under the optical stimulus in one bias direction. 
\\
\par  The observed phenomena of the light-emitting photodetection and light-enhanced electroluminescence in the simple MSM structures of CsPbBr$_3$ MWs underpins the unique structural complexity governed by mixed ionic-electronic conduction behaviour in the family of halide-perovskites. We expect that the cooperative action of charge carrier species (ions, photogenerated carriers) which leads to simultaneous light detection and emission will expand the applications towards multifuncitonal wearable and implantable devices needed for healthcare monitoring (e.g. pulse oximeters, smart contact lenses, optogenetics systems, neural monitoring) as well as will continue to develop in the advanced neuromorphic computing applications (which requires the memory function provided by ions) complemented by photodetection and optical switching (light-enhaced electroluminescence). The MSM halide-perovskite based devices will contribute to exploration of ionic-switching devices (e.g. ionic memories, ionic transistors), while the accurate analysis of this structure will help the researchers in the analysis of traditional optoelectronic-based devices dealing with materials with mixed ionic-electronic conduction behaviour.

\section*{Methods}

\textbf{Synthesis of CsPbBr$_3$ MWs}
\par We utilized temperature difference triggered growth method for the CsPbBr$_3$ MWs synthesis \cite{wang2018temperature}. The furnace (PZ 28-3TD High Temperature Titanium Hotplate and Program Regler PR5-3T) was used to controll the temperature during MWs growth. The CsPbBr$_3$ material was sublimated from the source substrate to the target substrate. On the glass source substrate we synthesised dense CsPbBr$_3$ microcystals according to the protocol \cite{pushkarev2018few}. The target substrate was a cleaned (5 min water and afterwards 5 min isopropanol places in ultrasonicat bath) sapphire. Two substrates were separated with air gap of 1 cm. The temperature of both substrates was controlled by the furnace temperature. The synthesis started at a furnace temperature temperature 350$^o$C. Then, the temperature rose up to 520$^o$C for 10 min and remained at this temperature for 10 min. As a result, the sublimated CsPbBr$_3$ material assembled on the sapphire substrates according to sapphire crystallographic planes. 
\textbf{Device fabrication}
\par Aerosol chemical vapour deposition was implemented to synthesize SWCNT \cite{tian2011controlled}. The as-fabricated SWCNT thin films contained both metallic and semiconducting SWCNT. The film thickness of 18 $\pm$ 5 nm was estimated by combined optical absorbance and ellipsometry measurements. Later, the films were cut with a blade into stripes of width 5 mm and length 7 mm. The stripes of single-walled carbon nanotube thin films were dry-transferred on the as-synthesized CsPbBr$_3$ microwire  sapphire substrates. Afterwards, 10 $\mu$l of isopropanol (IPA) was dropped on top of SWCNT and was annealed at 80 C$^{o}$ for 5 minutes. Then the ablation was performed with light conversion pharos femtosecond laser (200 fs pulse duration, wavelength 1030 nm). The light was focused through 50x lens (NIR Mitutoyo, NA = 0.65) while the xy position was altered with a programmable piezocontroller (Standa). In such geometry, where SWCNT stripe is located above the CsPbBr$_3$ microwire, the ablation fluence should exceed the threshold fluence for SWCNT material removal while at the same time should be less then two-photon ablation of CsPbBr$_3$ material. Hence, 1) pulse repetition rate was decreased down to 100 Hz and 2) the fluence level was controlled to be below 0.1 J/cm$^2$. The laser cut was performed in two steps. Firstly, the CsPbBr$_3$ microwire was chosen and the aforementioned ablation regime for SWCNT cut above the microwire was performed. Secondly, the fluence level was increased to ablate whole SWCNT stripe and to avoid other CsPbBr$_3$ microwire interconnects in the interelectrode space. 
\textbf{Optoelectronics measurements}
\par Optical measurements were performed using photoluminescence (PL) and electroluminescence (EL) spectroscopy in our custom-built experimental setup. For PL measurements the laser was focused on the samples using objective lens, the laser spot diameter was  30 $\mu m$. The excited signal (PL,EL) was collected by the same microscope objective. To excite samples we used 490 nm CW laser, to cut off excitation signal we used 500 nm long pass filter. The excited signal was measured by projecting the image plane of the objective onto the entrance slit of the imaging spectrometer (Princeton SP 2550), and detecting the light with a CCD detector (PyLoN 400BR eXcelon). The sample was mounted in an ultra-low-vibration closed-cycle helium cryostat (Advanced Research Systems, DMX-20-OM), sample holder was augmented with electrical pins for optoelectronics measurements. The optical measurements were performed on the same optical setup for AC and DC electrical measurements. 

\par We used two different electrical schemes for dynamic (AC) and current-voltage curves (DC) measurements, which connected to the sample via the same port in cryostat. To measure current-voltage curves in DC regime we used Keithley 2401 Source meter. Instrument applied voltage of bias of 20 ms pulse, and the obtained current signal which was averaged through 1 second period. The scan rate was fixed to be 0.25 V per second. 
Optical signal collected the same way and the illumination intensity of 490 nm CW laser was tuned through the optical filter ranged from 0.06 to 60 mW cm$^{-2}$. To electrically excite our sample in AC, we used wave generator (Keysight 33600A), which allowed us to vary duty cycle, voltage amplitude and pulse repetition frequency independently. The current signal was collected through the shunt resistance $R_{shunt}$ = 8.2 kOm on the Oscilloscope (Keysight DSOX6004A). In case of 10 Hz measurements, light illumination was chopped accordingly pulse repetition frequency, so that illumination turned-on on the scale of applied voltage pulse. The electrical measurements coordinated in time with the spectrometer, so that the device response (e.g. EL,PD) was measured simultaneously. 

\bibliography{sample}

\section*{Acknowledgements}
A.A.M. acknowledges Ivan Tzibizov and Dr. Fedor Benimetsky for the discussion on cryostat measurements, and professor Prof. Anvar Zakhidov for valuable discussions. The authors are thankful to the Priority 2030 Federal Academic Leadership Program and acknowledge Russian Science Foundation (Project Number 19-73-30023). A.G.N. acknowledges Russian Science Foundation (Project Number 22-13-00436) for supporting SWCNT synthesis part. The results were partially obtained on the equipment of the ITMO Core Facility Center “Nanotechnologies”.

\section*{Author contributions statement}
A.A.M. observed LEPD regime and originated the idea. A.G.N. synthesized single-walled carbon nanotube thin films. A.A.M. fabricated the device. A.A.M. and A.P.P. discussed the device design and operation. S.A.K., A.P.P. and A.A.M. contributed to XPS measurements and analysis. A.A.M. and V.I.K. performed optoelectronic measurements. M.A.B. performed SEM and EDX measurements. A.A.M., V.I.K., A.P.P. and S.V.M. analyzed the data. S.V.M. supervised project. A.A.M. wrote the original draft. All authors reviewed and edited manuscript. All authors contributed to the discussions and commented on the paper.

\section*{Additional information}
The authors declare no competing interests.

\end{document}


\title{

Supplementary material: Halide Perovskite Light-Emitting Photodetector
}

\author[1,*]{A.A.Marunchenko}
\author[1]{V.I.Kondratiev}
\author[1]{A.P.Pushkarev}
\author[1]{S.A.Khubezhov}
\author[1]{M.A.Baranov}
\author[2]{A.G.Nasibulin}
\author[1,3,4*]{S.V.Makarov}

\affil[1]{ITMO University, School of Physics and Engineering, St. Petersburg, 197101, Russian Federation}
\affil[2]{Skolkovo Institute of Science and Technology, 30/1 Bolshoy Boulevard, 121205 Moscow, Russian Federation}
\affil[3]{Harbin Engineering University, Harbin 150001, Heilongjiang, China}
\affil[4]{Qingdao Innovation and Development Center of Harbin Engineering University, Qingdao 266000, Shandong, China}
\affil[*]{Corresponding author}

\makeatletter
\renewcommand{\@maketitle}{%
{%
\thispagestyle{empty}%
\vskip-36pt%
{\raggedright\sffamily\bfseries\fontsize{20}{25}\selectfont \@title\par}%
\vskip10pt
{\raggedright\sffamily\fontsize{12}{16}\selectfont  \@author\par}
\vskip25pt%
}%
}%
\makeatother
\maketitle

\begin{figure}[t!]
\centering
\center{\includegraphics[width=1.0\linewidth]{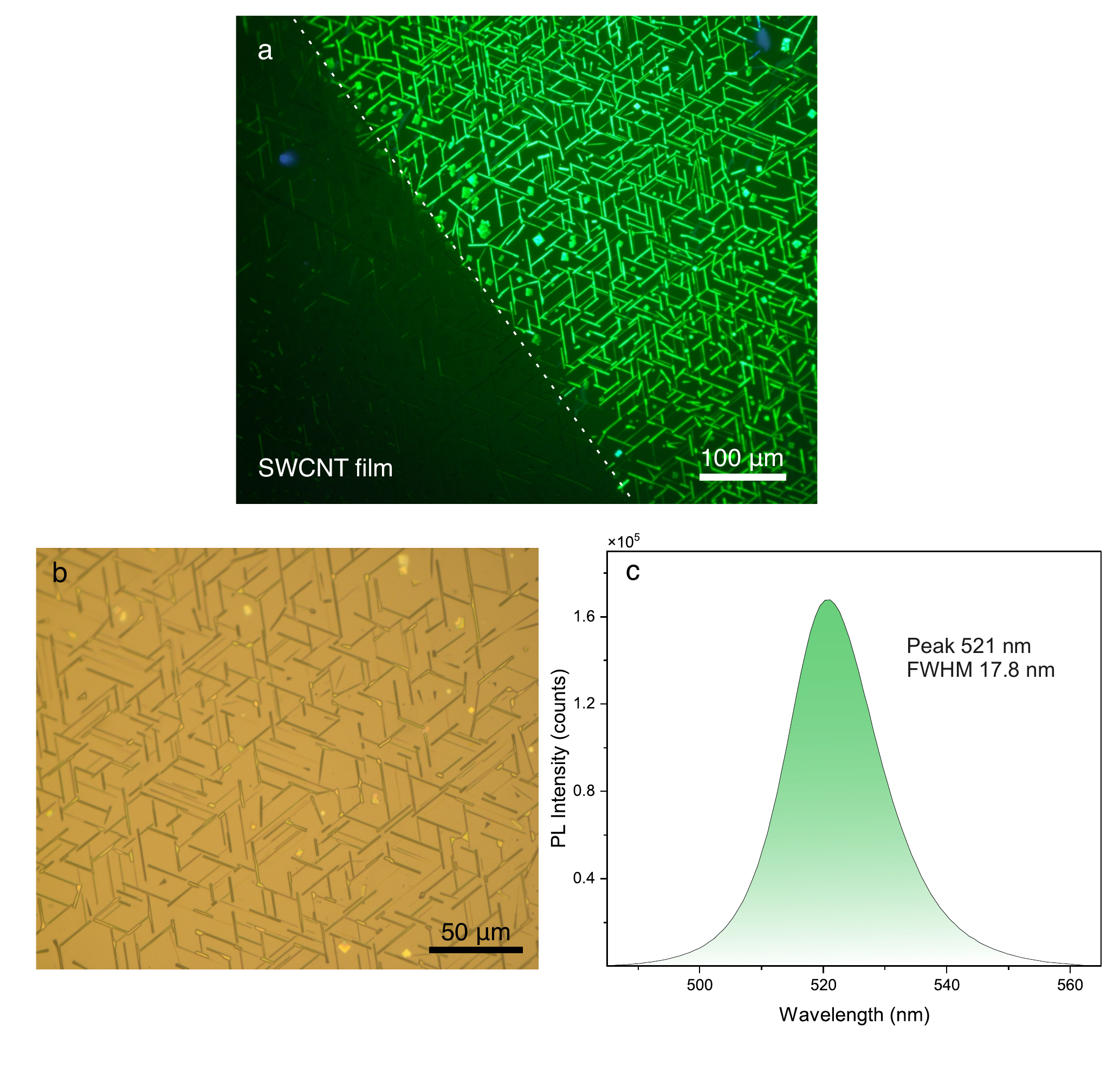}}
\caption{ | \textbf{Optical image and PL of as-synthesized CsPbBr$_3$ microwires on a sapphire substrate}. a) PL image of big area of patterned CsPbBr$_3$ microwires (MWs) grown on the sapphire substrate. Part of the wires covered with SWCNT thin film. b) Optical image of CsPbBr$_3$ MWs. c) PL spectrum of the set of CsPbBr$_3$ MWs at room temperature. 
}
\label{fig:concept}
\end{figure}

\clearpage

\begin{figure}[t!]
\centering
\center{\includegraphics[width=1.0\linewidth]{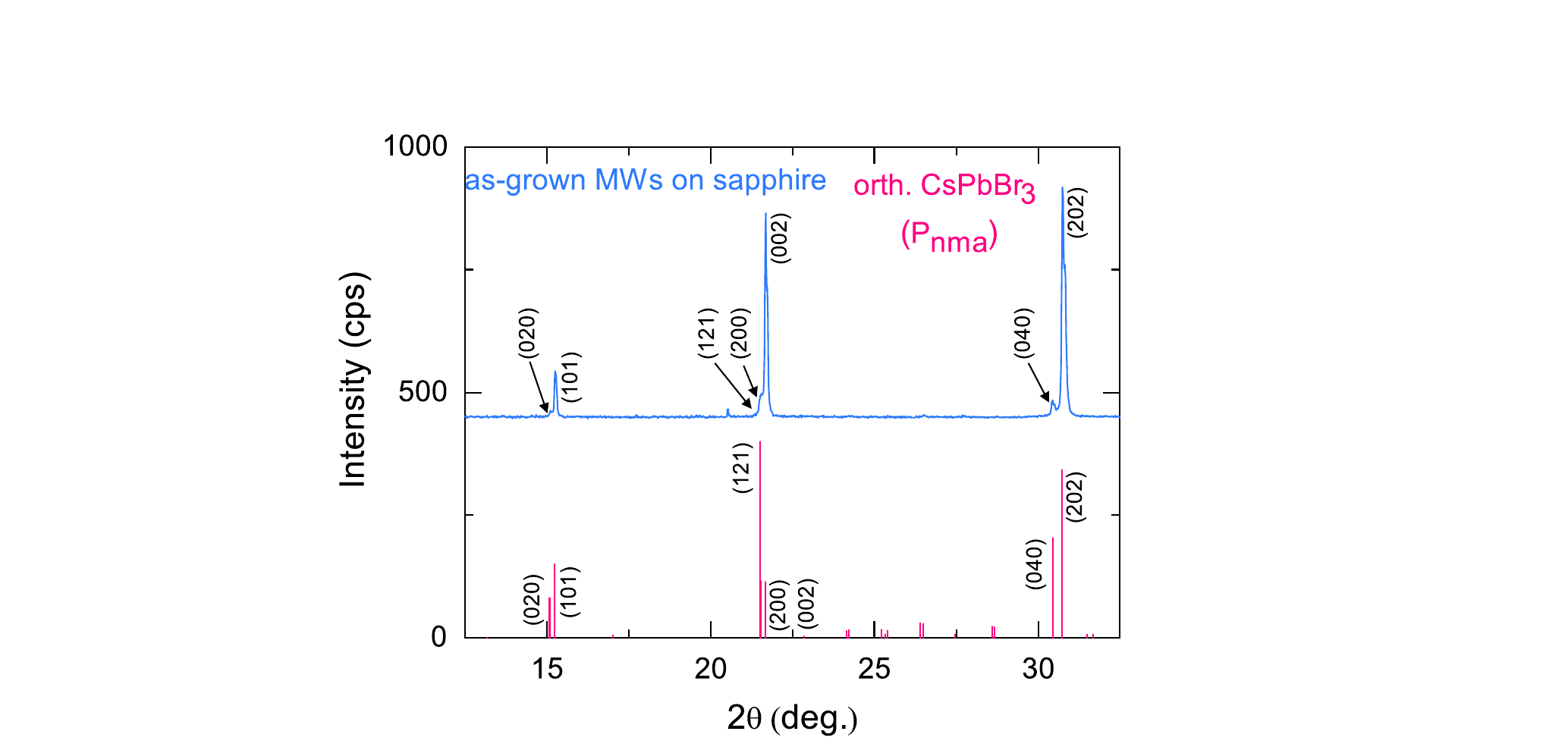}}
\caption{ | \textbf{XRD spectrum of as-grown CsPbBr$_3$ MWs on sapphire}. XRD spectrum was collected with Rigaku Ultima IV X-ray diffractometer. The peaks of as-grown CsPbBr$_3$ MWs correlate to orthorhombic crystal structure (space group P$_{nma}$).
}
\label{fig:concept}
\end{figure}

\clearpage

\begin{figure}[t!]
\centering
\center{\includegraphics[width=1.0\linewidth]{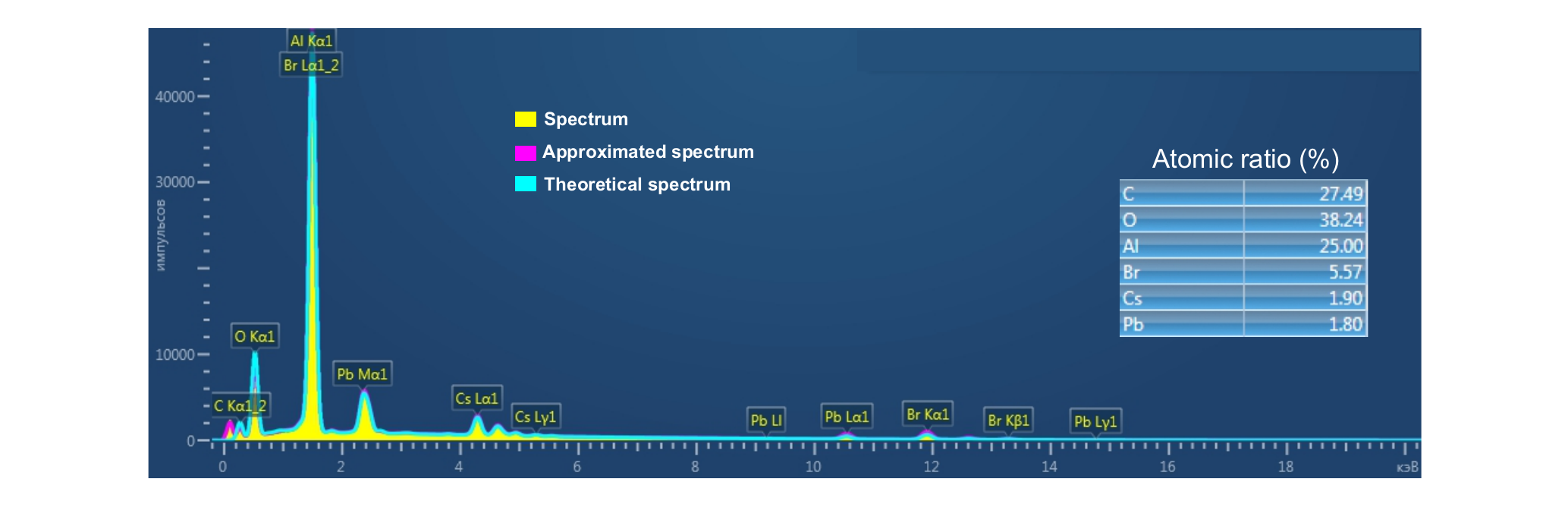}}
\caption{ | \textbf{EDX spectrum}. The spectrum shows stoichiometric atomic ratio between Cs:Pb:Br elements equal to $\sim$ 1:1:3. 
}
\label{fig:concept}
\end{figure}

\clearpage

\newpage
\begin{figure}[t!]
\centering
\center{\includegraphics[width=1.0\linewidth]{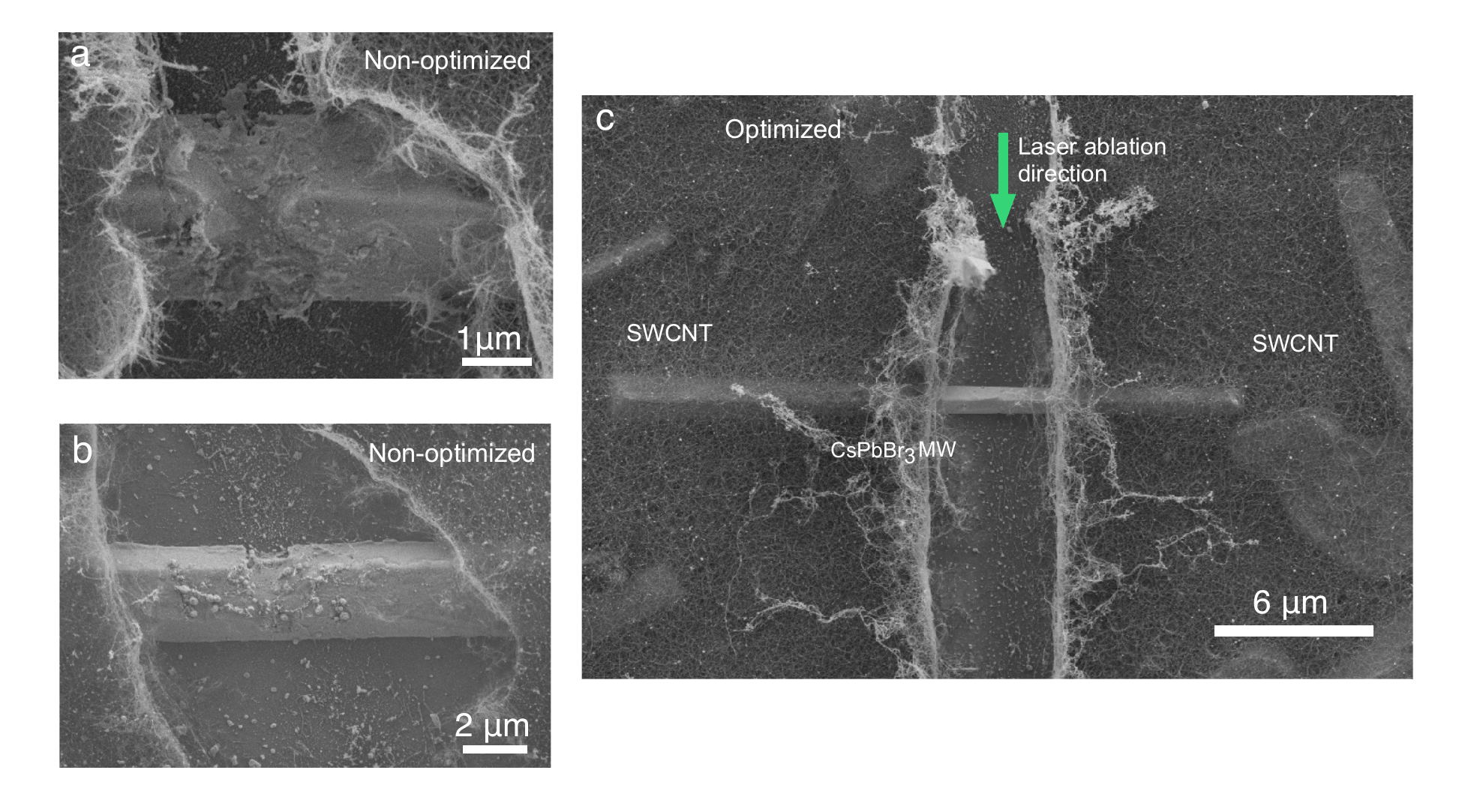}}
\caption{ | \textbf{SEM image of CsPbBr$_3$ MWs after ablation.} SEM images performed by Zeiss MERLIN system. a,b) Typical CsPbBr$_3$ MWs between SWCNT thin film when the ablation conditions are not optimized (e.g. fluence, scan rate, pulse repetition frequency). c) Example of MW at optimized ablation conditions. The laser ablation direction is visualized. The thin thickness of SWCNT film allows to visualize CsPbBr$_3$ crystals underneath. 
}
\label{fig:concept}
\end{figure}

\clearpage

\newpage \par \textbf{Analysis of chemical content of perovskite microcrystals}
\par For the identification of a chemical content of perovskite crystals fabricated by temperature difference triggered growth method X-ray photoelectron spectroscopy (XPS) depth profiling of a single CsPbBr$_3$ microplate with lateral dimensions about 10$\times$10 $\mu$m is conducted. The microplate is grown on Si substrate to diminish its charging during the XPS analysis. XPS spectra are measured at 2.4$\times$10$^{-9}$ mbar pressure on a spectrometer (Thermo Scientific K-Alpha) equipped with an Al source of monochromatic K-alpha X-ray ($h\nu$ = 1486.6 eV). An energy scale is adjusted
in accordance with the Au 4f = 84 eV, Ag 3d = 368.2 eV and Cu 2p = 932.6 eV lines of standard samples. To obtain high-resolution (0.1 eV) spectra, the x-ray beam is focused onto a 30 $\mu$m spot at a constant pass energy of 20 eV and the number of scans of 10 is chosen. The surface etching is done with Ar$^+$ ions accelerated in potential of 500 V. The rate of etching is 0.2 nm s$^{-1}$.

\par The depth profiling shows a 1 nm thin contaminating layer over perovskite material (Fig. S5a). This layer contains carbon derivatives absorbed by the surface of microcrystal from atmosphere that exhibit a signal peaked at 284.7 eV. Note all the spectra contain Si 2p peak because the lateral dimensions of the studied microplate are smaller than diameter of the x-ray spot (Fig. S5b). A peak at 103.1 eV is assigned to silicon oxide layer which completely disappears when the etch depth exceeds 8.4 nm. Along with that, a peak at 99.2 eV corresponding to pure Si saturates at the depth of 8.4 nm~\cite{https://doi.org/10.18434/t4t88k}. Spectral position of Cs 3d (725 and 739 eV) and Br 3p (182.6 and 189.3 eV) doublets is typical for CsPbBr$_3$ (Fig. S5c,d)~\cite{liashenko2019electronic}. The intensity of these doublets alters until the etch depth exceeds 4.8 nm and thereafter takes almost stationary value. The latter implies that bulk perovskite is most likely covered with a thin layer of some other material. Such behavior is complemented by the depth-dependent line shape of Pb 4f and O 1s peaks. Looking at Pb 4f doublet collected from the microcrystal surface, one can see its complex line shape (Fig. S5e). Each of the doublet lines consists of 3 peaks associated with Pb$-$Pb (136.78 and 141.68 eV), Pb$-$O (138.2 and 143.1 eV), and Pb$-$Br (139.0 and 143.8 eV) bonds corresponding to metallic lead, lead(II) oxide, and cesium lead(II) tribromide, respectively~\cite{payne2005lead,liashenko2019electronic}. Expectedly, O 1s signal demonstrates a complex line shape as well (Fig. S5f). It can be approximated by a sum of 5 peaks belonging to Pb$-$O (530.0 eV), Pb$-$Br (138.9 eV), C$=$O (531.5 eV) , C$-$O (532.27 eV), OH and V$^{+2}_O$ (531.0 eV), and Si$-$O (532.8 eV) bonds. 

\begin{figure}[t!]
\centering
\center{\includegraphics[width=1.0\linewidth]{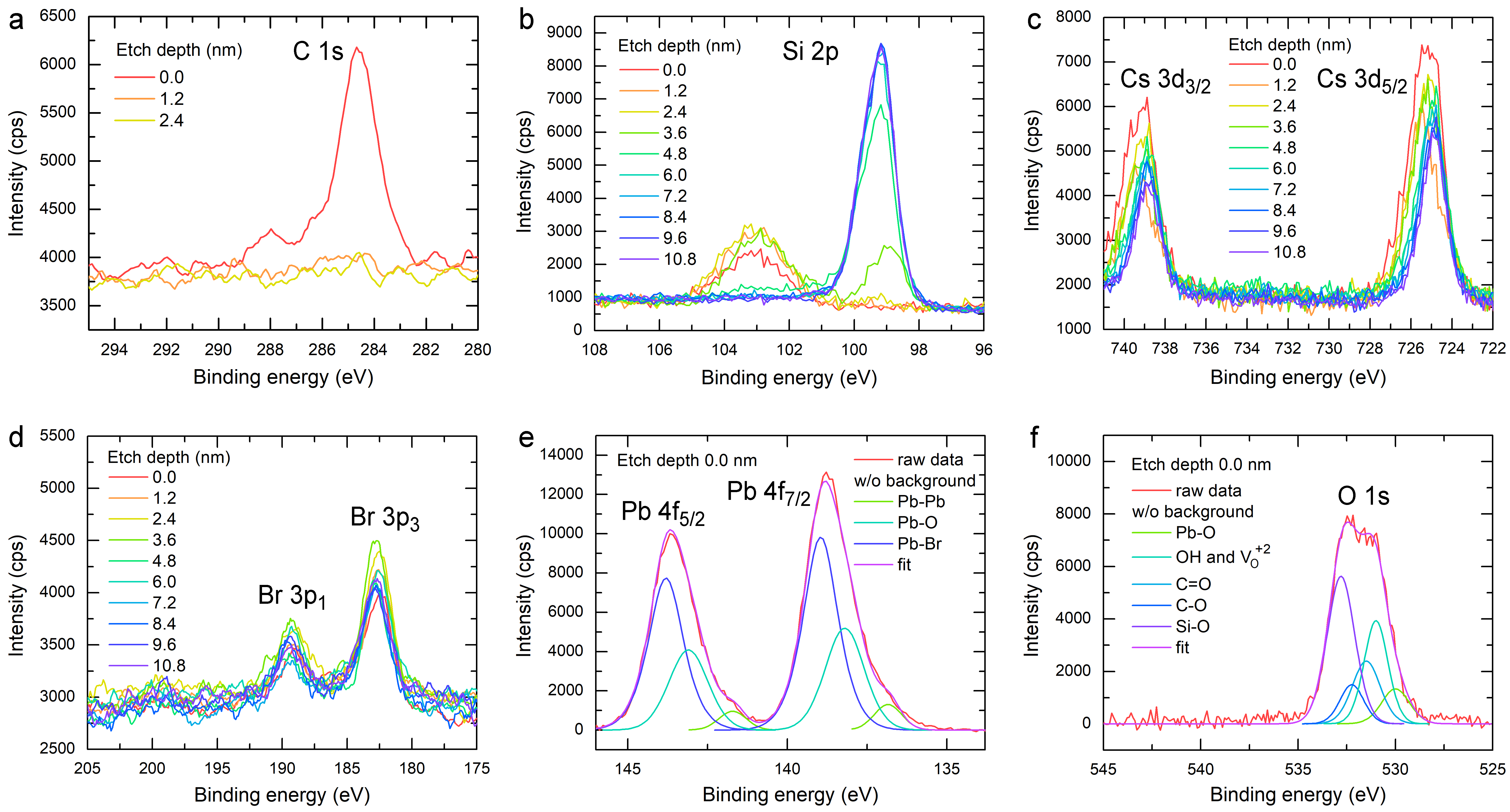}}
\caption{(a-d) High-resolution XPS profiling data for C 1s (a), Si 2p (b), Cs 3d (c), and Br 3p (d) signals. (e,f) Spectra from the surface of microcrystal showing a complex line shape of Pb 4f (e) and O 1s (d) signals. Both spectra contain lines assigned to Pb$-$O bond that indicates the presence of lead(II) oxide on top of the microcrystal.}
\label{fig:xps}
\end{figure}
\begin{figure}[t!]
\centering
\center{\includegraphics[width=1.0\linewidth]{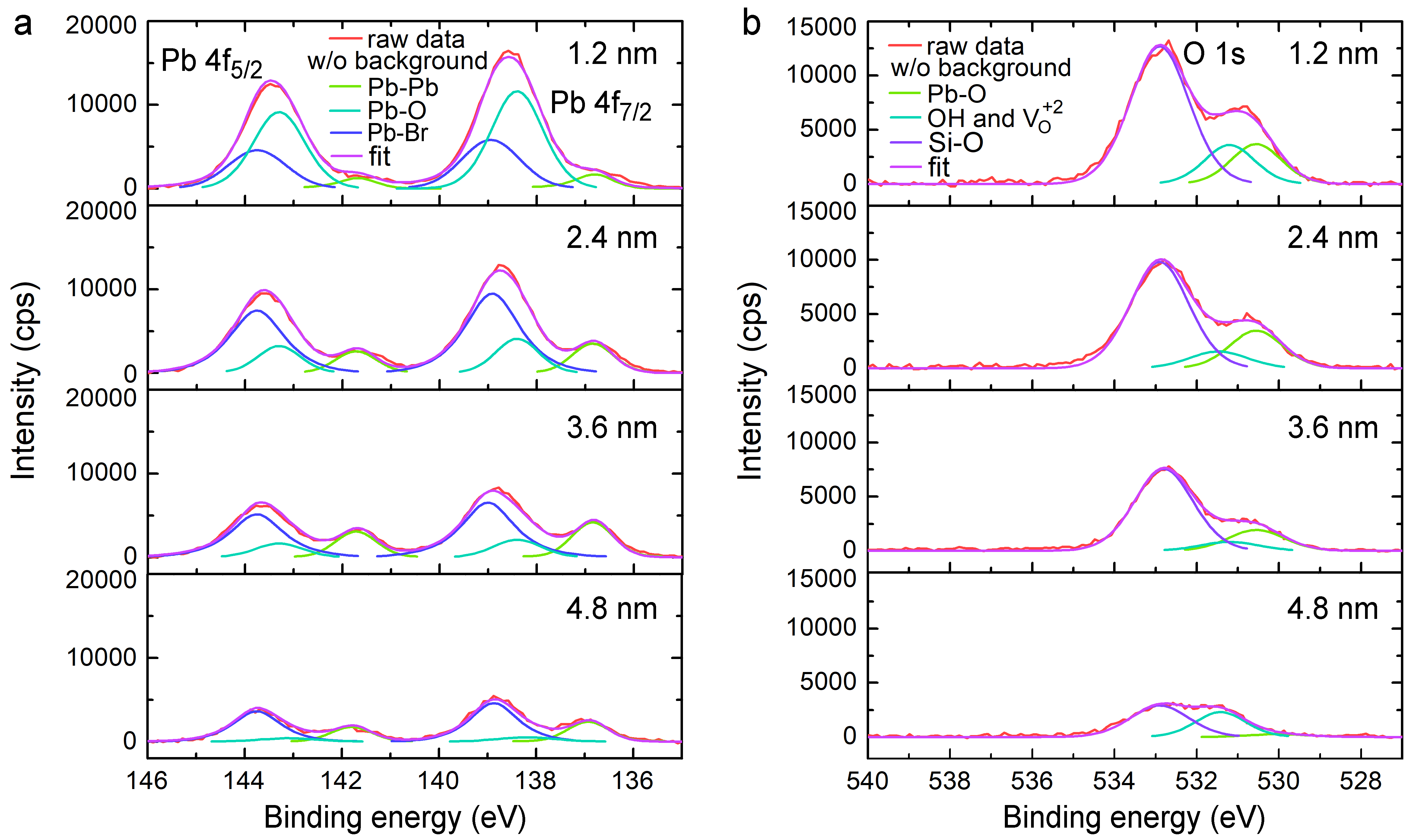}}
\caption{(a,b) Etch depth-dependent evolution of Pb 4f (a) and O 1s (b) XPS signals. Both spectral dynamics reveal a gradual decrease in the intensity of peaks corresponding to Pb$-$O bond as compared to the intensity of other lines.}
\label{fig:xps}
\end{figure}

\begin{figure}[t!]
\centering
\center{\includegraphics[width=1.0\linewidth]{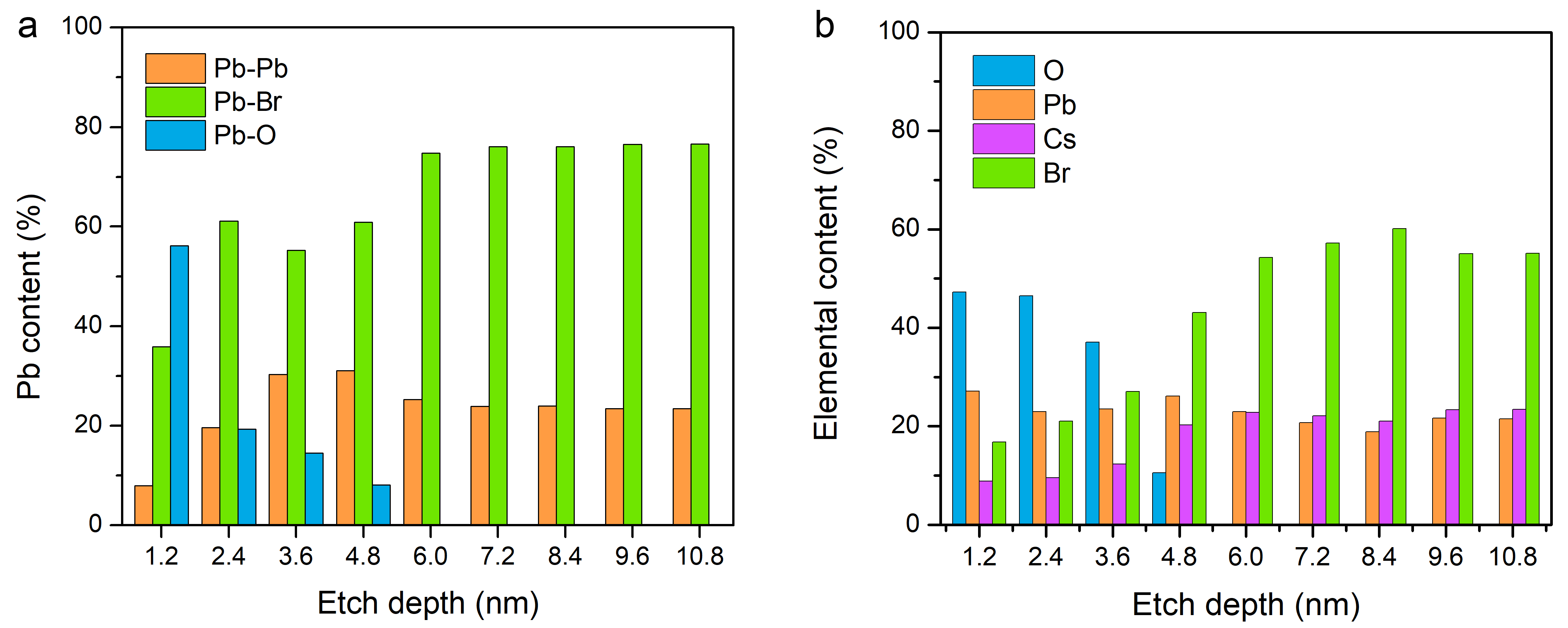}}
\caption{(a,b) Etch depth-dependent concentration of lead species (a) and all the chemical elements (b) constituting CsPbBr$_3$ microcrystal covered with a thin PbO layer.}
\label{fig:xps}
\end{figure}
\par Etching the surface gives a gradual decrease in Pb$-$O lines intensity and simultaneous increase in Pb$-$Br and Pb$-$Pb ones (Fig. S6a,b). Although the contrary dynamics of Pb$-$O and Pb$-$Br signals is obviously related with transition from naturally formed PbO layer at the surface to bulk CsPbBr$_3$ that, in turn, explains aforementioned dynamics established for Cs 3d and Br 3p lines, evolution of the Pb$-$Pb signal needs to be clarified. We think the reason for that could be the reduction of Pb$^{2+}$ caused by removing O and Br atoms from PbO and CsPbBr$_3$ crystal lattices, respectively, during their bombardment with Ar$^+$ ions. Since a covalent polar Pb$-$O bond is stronger than ionic Pb$-$Br one, the reduction of lead(II) is a preferable process for the latter as compared to the former.

Based on the depth profiling data the relative content of different lead species is estimated (Fig. S7a). It is established, that 5 nm layer of PbO is formed on the surface of the perovskite microplate. Note the boundary between PbO and CsPbBr$_3$ can not be defined precisely because a penetration depth of x-ray is about 4.2$\pm$1.2 nm. Furthermore, by subtracting Si$-$O, OH, and V$^{+2}_O$ signals from the spectra presented in Figure S6b we can employ exclusively Pb$-$O signal and the rest of the collected data (Fig. S5c,d; Fig. S6b) for rigorous evaluation of depth-dependent relative concentration of O, Cs, Pb, and Br in the CsPbBr$_3$ microplate covered with thin lead(II) oxide layer (Fig. S3b). Calculations are done according to the following formulae:
\begin{equation}
  N_{i}^{element} = \frac{S^{norm}_{i}}{\Sigma^{n}_{i=1}~S^{norm}_{i}}, ~~S^{norm}_{i} = \frac{{S^{peak}_{i}}}{F_iT_iC_i}, 
\end{equation}
where $N_{i}^{element}$ is the concentration of a certain ($i$) chemical element, $S_{i}^{norm}$ is the normalized area of its peak, $S_{i}^{peak}$ – area of its peak, $F_i$ - sensitivity factor for the element, $T_i$ – transmission function, $C_i$ - the decay length. The values for $F_i$, $T_i$, and $C_i$ were taken elsewhere~\cite{scofield1976hartree}. As a result, XPS depth profiling of the studied sample confirms the existence of 4.2$\pm$1.2 nm layer of naturally formed lead(II) oxide on top of the CsPbBr$_3$ microcrystal. It should be pointed out that in the main body we omit the layer contaminated with carbon derivatives for clarity of presentation of the data. For this reason the point 1.2 nm in Figure S7b corresponds to the point 0 nm in Figure 1c.

\clearpage

\newpage

\begin{figure}[t!]
\center{\includegraphics[width=1.0\linewidth]{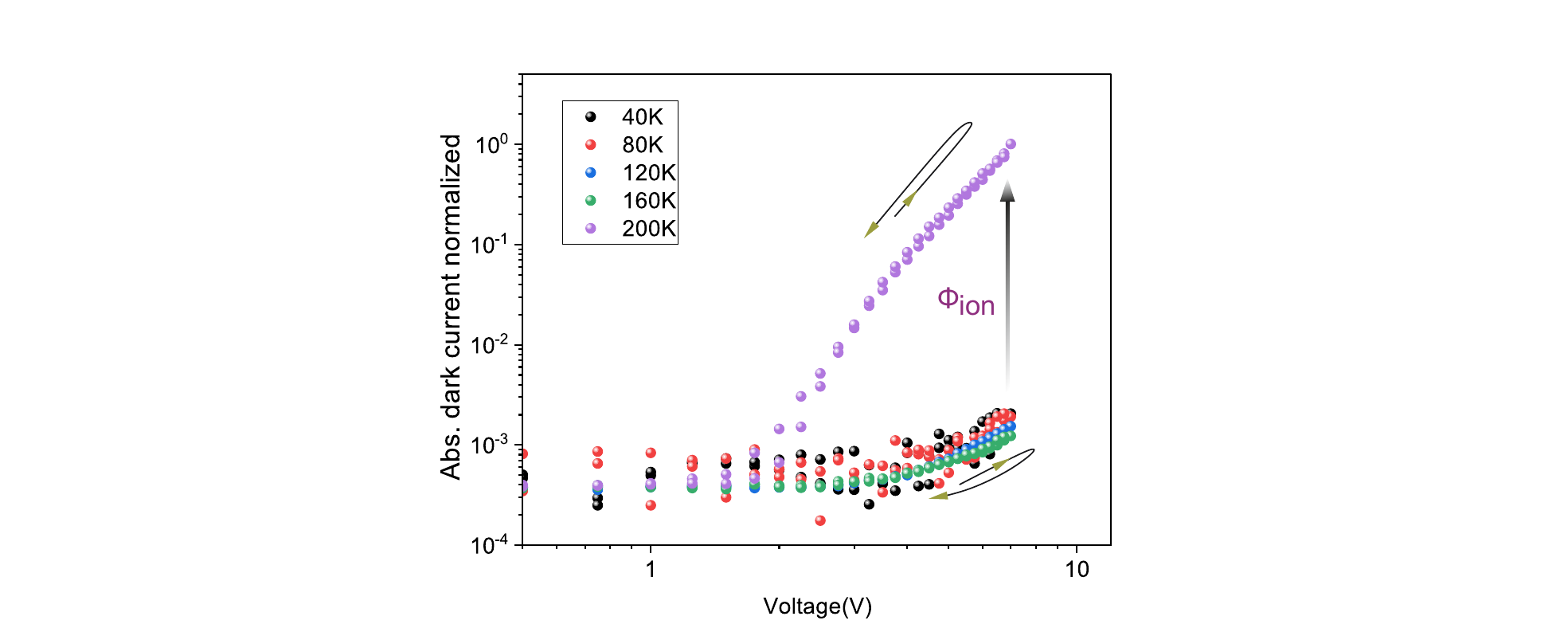}}
\caption{ | \textbf{Dark current gating by ions.} Absolute normalized dark current values in temperature range 40 K $-$ 200 K. The significant dark current modification at 200 K reflects the dark current gate by mobile perovskite ions $\Phi_{ion}$. In addition, the direction of current hysteresis changes at 200 K. The temperature dependence of dark current below 200 K does not follow the dependence of thermionic current emission theory.}
\label{fig:concept}
\end{figure}

\clearpage
\newpage

\begin{figure}[t!]
\center{\includegraphics[width=1.0\linewidth]{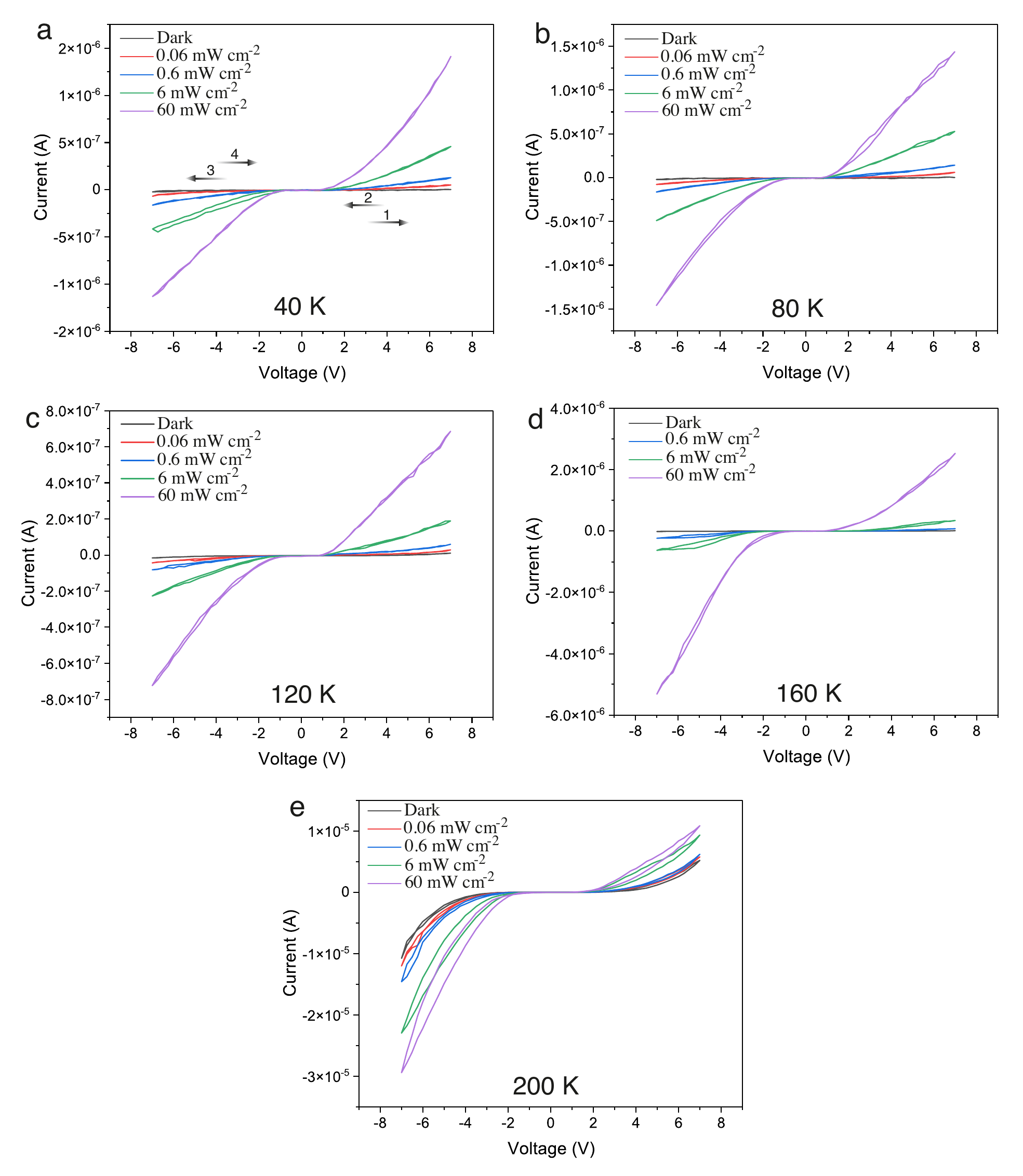}}
\caption{ | \textbf{I-V in dark and under illumination for 40 $-$ 200 K temperature range.}  a-e) Current-voltage curves for temperatures 40 K, 80 K, 120 K, 160 K, 200 K respectively in both positive and negative applied bias directions at a scan rate of 0.25 V per second. The direction of the scan rates is the same at all temperatures (0 V | 7 V | 0 V | -7 V | 0 V) and is visualized in a). In all cases the photocurrent gradually increases under rising illumination intensity from 0.06 mW cm$^{-2}$ to 60 mW cm$^{-2}$.}
\label{fig:concept}
\end{figure}

\clearpage
\newpage

\begin{figure}[t!]
\center{\includegraphics[width=1.0\linewidth]{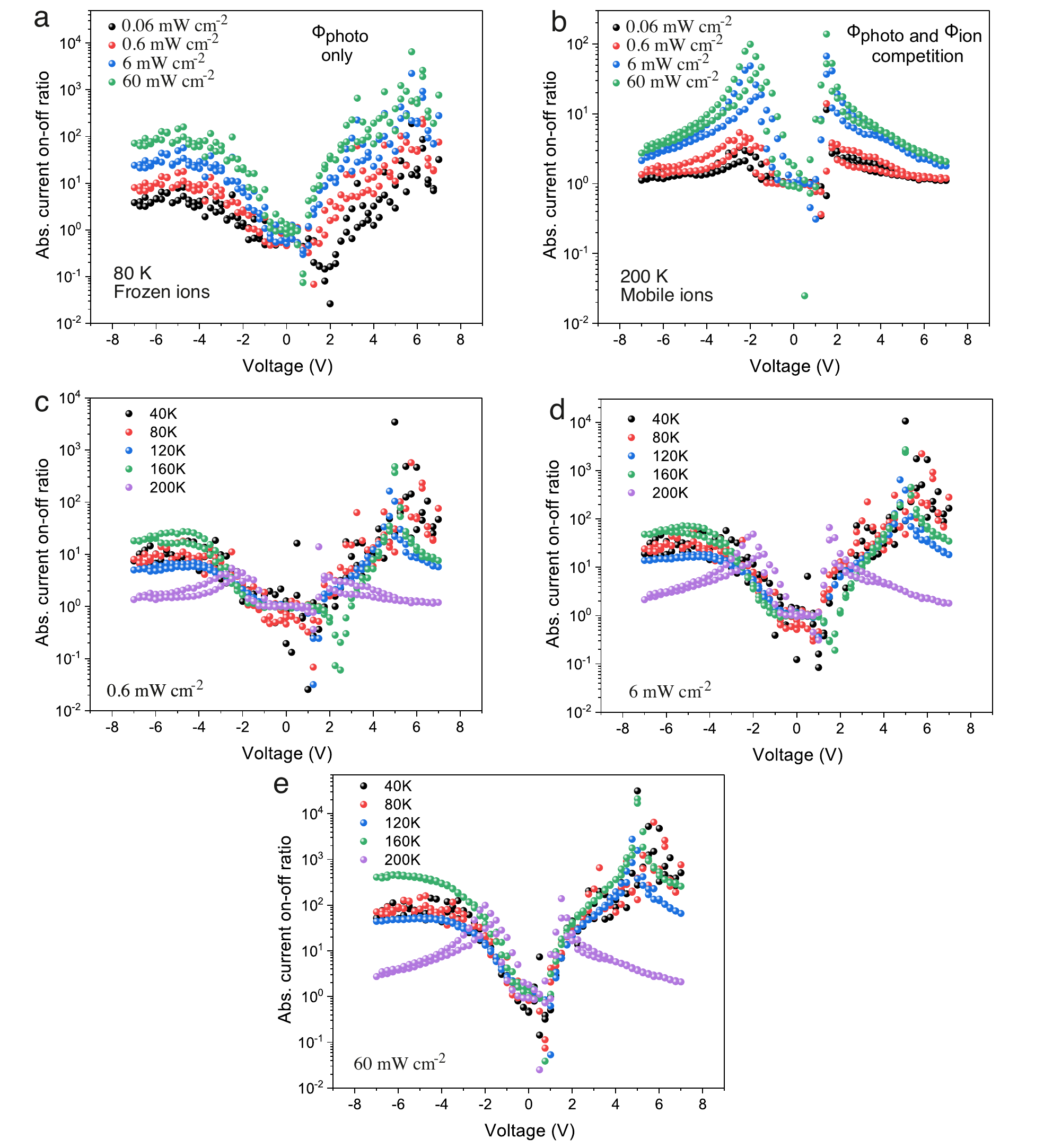}}
\caption{ | \textbf{Absolute current on-off ratio ($|I_{photo}/I_{dark}|$) at varied illuminating conditions and temperature.} a,b) The principle comparison between the cases of a) frozen ions  and b) mobile ions. The appearance of mobile ions lead to fast current on-off ratio decrease at higher voltages due to the competition between the $\Phi_{photo}$ and $\Phi_{ion}$ current gates. Hence, in halide-perovksite MSM structure, the presence of mobile ions can be checked by current on-off ratio graph. c-e) Absolute current on-off ratio values at different temperatures in range 40 K $-$ 200 K and illumination intensity  c) 0.6 mW cm$^{-2}$, d) 6 mW cm$^{-2}$ and e) 60 mW cm$^{-2}$.}
\label{fig:concept}
\end{figure}

\clearpage
\newpage

\begin{figure}[t!]
\center{\includegraphics[width=1.0\linewidth]{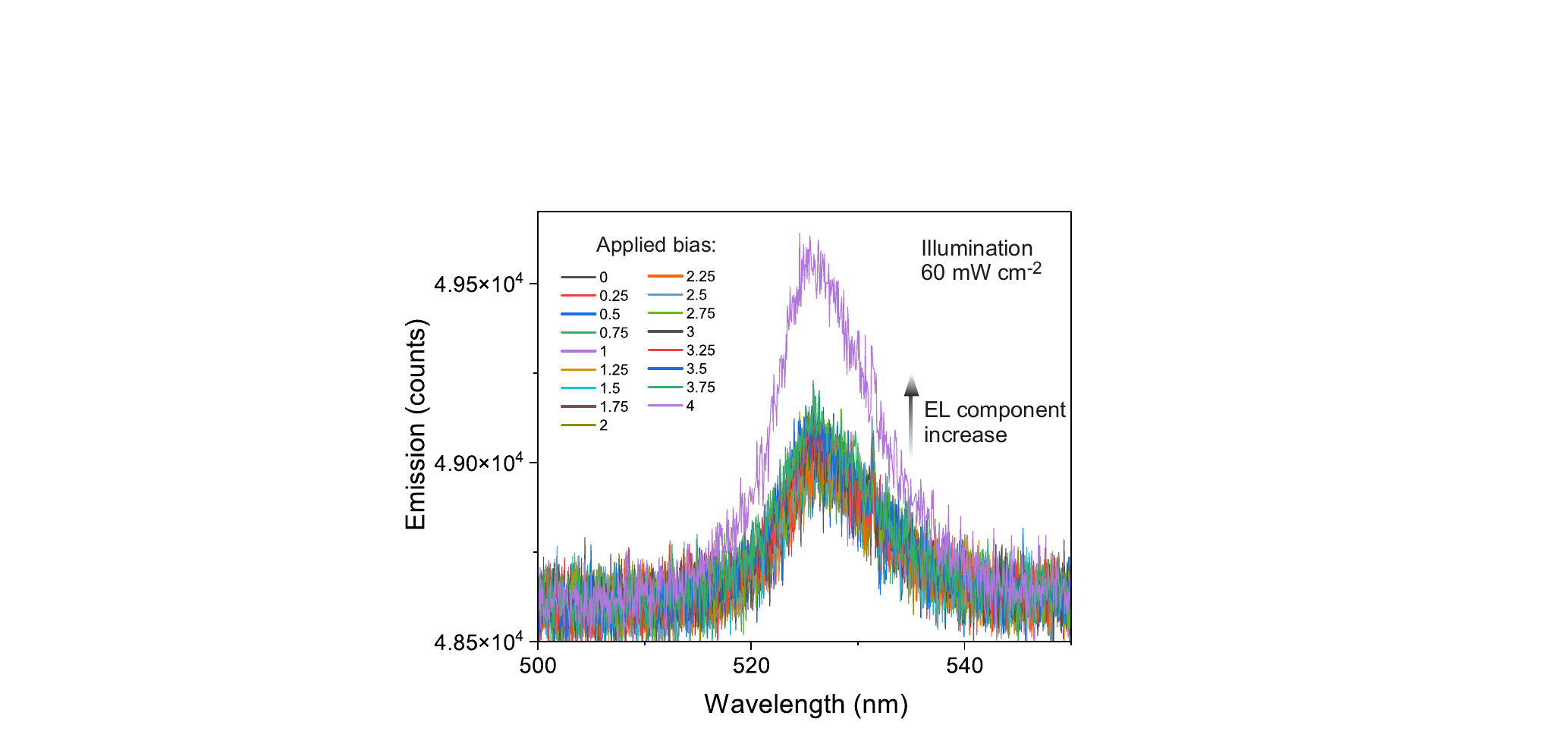}}
\caption{ | \textbf{Position of PL(EL) under bias.} Under illumination 60 mW cm$^{-2}$ the total spectrum (EL+PL) is shown. At the applied bias 4 V, the EL component is visually separated. The position of separate EL and PL peaks can not be resolved.}
\label{fig:concept}
\end{figure}

\clearpage
\newpage

\begin{figure}[t!]
\center{\includegraphics[width=1.0\linewidth]{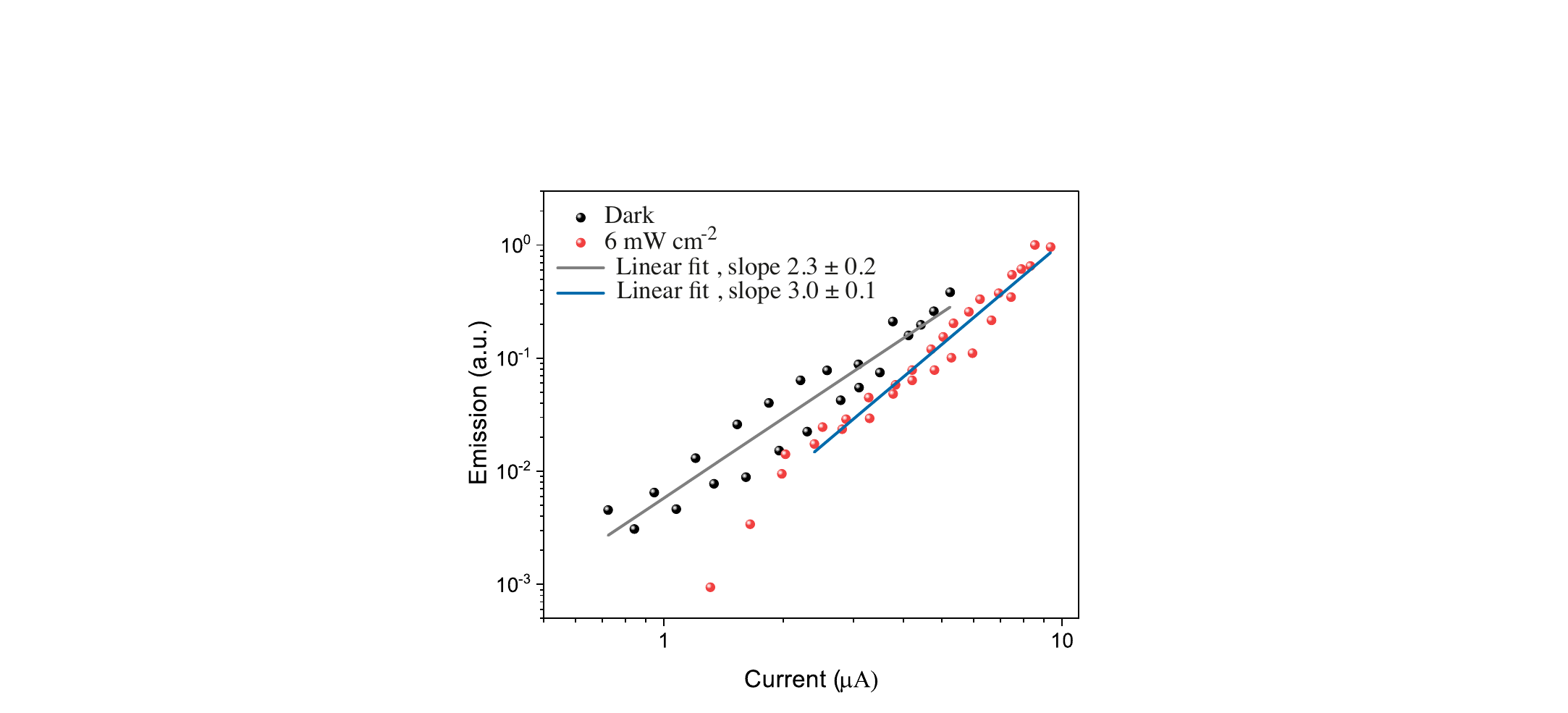}}
\caption{ | \textbf{Log-log plot of emission intensity A versus current I flowing in the CsPbBr$_3$ MW.} For the dark fit, A $\sim$ I$^{2}$ (2.3$\pm$0.2) is in accordance with ABC model for perovskite recombination with trap-state dominant recombination term. For the fitting in illuminated case, the dependence on current A $\sim$ I$^{k}$, where k>2 reflects change in carrier dynamics. The emission at the same current value for illuminated case is smaller than in the dark case which means that only part of the photocurrent contribute to the radiative recombination.}
\label{fig:concept}
\end{figure}

\clearpage
\newpage

\begin{figure}[t!]
\center{\includegraphics[width=1.0\linewidth]{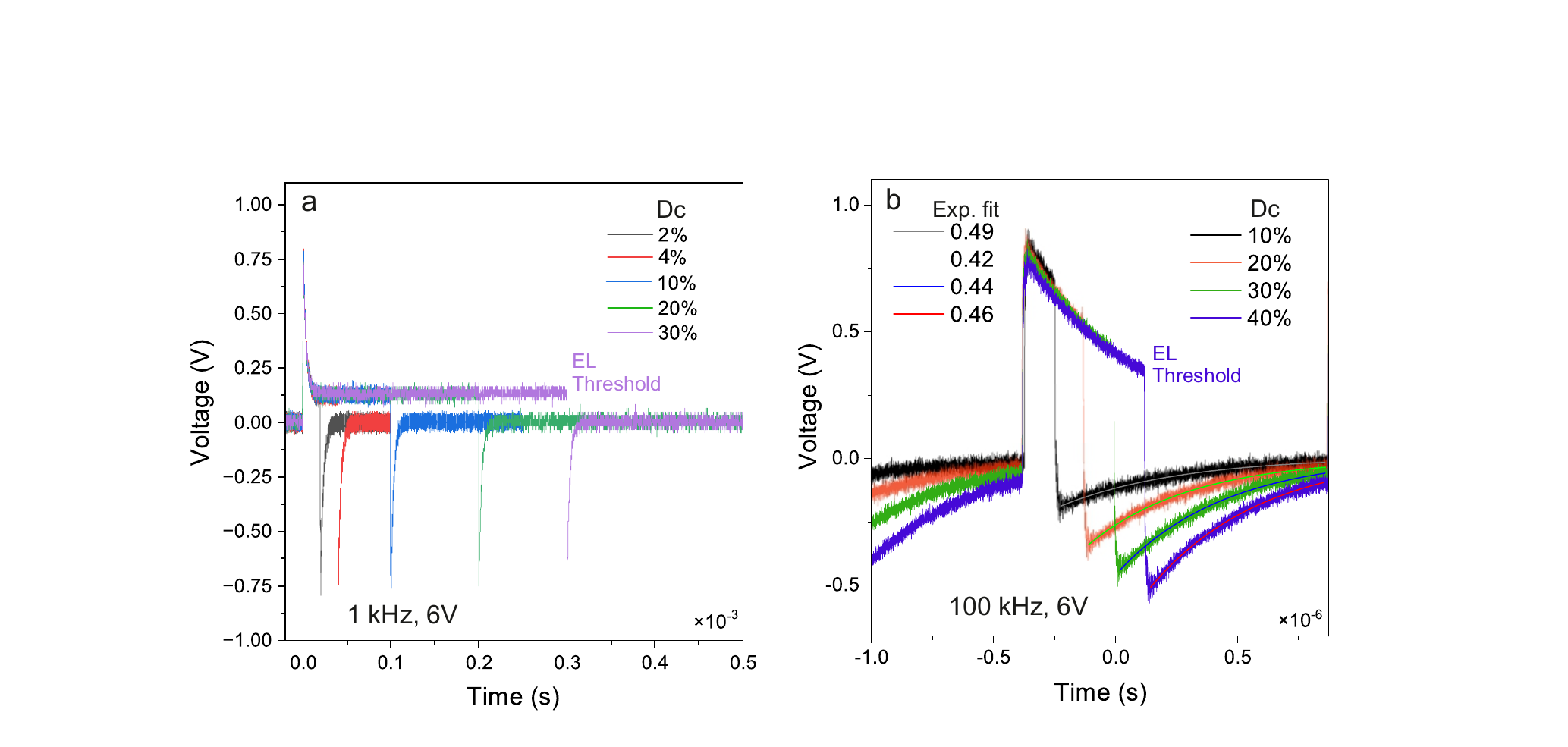}}
\caption{|\textbf{EL threshold at higher frequency and 6 V applied bias} a) At 1 kHz, the EL threshold is achieved at 30$\%$ Dc similar to low-frequency case (10 Hz). b) The EL threshold achieved at 40$\%$ Dc for 100 kHz. The sharp voltage spikes (a-b) can be attributed to the capacitance of PbO. At 100 kHz spikes are resolved in microsecond time domain at 6 V of applied bias. The negative spikes at the end of each pulse reflect the capacitive discharge. The monoexponential fits of capacitive discharge at different Dc are also shown.}
\label{fig:concept}
\end{figure}

\clearpage
\newpage

\bibliography{sample}